# Physics of Propagation in Left-Handed Guided Wave Structures at Microwave and Millimeter Wave Frequencies

Clifford M. Krowne*

Microwave Technology Branch, Electronics Science & Technology Division, Naval Research Laboratory, Washington, D.C. 20375-5347

**ABSTRACT**

A microstrip configuration has been loaded with a left-handed medium substrate and studied regarding its dispersion diagrams over the microwave and millimeter wave frequency bands for a number of different modal solutions. Ab initio calculations have been accomplished self-consistently with a computer code using a full-wave integral equation numerical method based upon a Green's function employing appropriate boundary conditions. Bands of both propagating and evanescent behavior have been discovered in some of the modes. For such modes, electromagnetic field plots in the cross-sectional dimension have been made. New electric field line and magnetic circulation patterns are discovered.

* Email address: krowne@chrisco.nrl.navy.mil

Tremendous interest in the last few years has occurred with the experimental realization of macroscopic demonstrations of left-handed media, predicted or at least suggested in the literature several decades ago [1]. Attention has followed on the focusing characteristics of left-hand media, with appropriate arrangements to accomplish such behavior, as shown by literature publications [2-8]. But no attention has been directed toward what left-handed media could accomplish in propagating devices used in integrated circuit configurations. Much of the work to date has looked at macroscopic realizations, which may be amenable in the future with current efforts on metamaterials, to advancing microwave integrated circuit component technology utilizing left-handed media.

There may be substantial interest in understanding the effects of left-handed media in guided wave structures since advances in integrated circuit technology, in passive components, control components, and active devices has increasingly been utilizing layers and arrangements of many differing materials. From heterostructures in active devices to complex materials like chiral, ferroelectric and ferromagnetic materials, in passive structures and control components, this trend has been increasing. Efforts in metamaterials is sure to further this trend.

We have found that modifying the anisotropic properties of ferroelectric permittivity tensors [9, 10] subject to static electric bias field [11] results in varying propagation behavior. Similarly, we have found layers of ferromagnetic materials to undergo significant permeability tensor [12] changes with change of static bias field magnitude and direction, and magnetization. Examination of the electromagnetic field distribution which comes about from the nonreciprocity embedded in the asymmetric permeability tensors, demonstrates the ability to modify the intensity and field line directions of the electric and magnetic fields.

In this Letter, we report on the new physics associated with altering the electromagnetic fields of guided wave propagating structures using left-handed media at microwaves and millimeter waves. Here we will only address the use of left-handed media with permittivity and permeability tensors reducing to scalars $\varepsilon(\omega)$ and $\mu(\omega)$. Clearly, both by individual unit cell construction and repetitive cells in favored directions, anisotropy can be introduced. As seen already in the literature, obtaining the characteristics of left-handed behavior, i.e. where $\text{Re}[\varepsilon(\omega)] < 0$ and $\text{Re}[\mu(\omega)] < 0$ simultaneously, may be a narrow band phenomenon [13 - 15]. (There are some indications or thoughts that non-resonant phenomena could eventually allow broader band devices [16], although this idea has come from focusing studies.) Thus the frequency region $\Delta\omega$ over which the desired behavior occurs may almost be viewed as being created by a Dirac delta selection function operating on the structure characteristics about a particular $\omega$. So in order to study what a left-handed substrate would do in a certain configuration at a particular frequency, we merely need to set $\text{Re}[\varepsilon(\omega)] = -\varepsilon_r$ and $\text{Re}[\mu(\omega)] = -\mu_r$ where $\varepsilon_r$, $\mu_r$ = real positive constants. Sweeping the frequency over a range for these settings will yield the interaction of the guiding structure on the left-handed medium, giving the fundamental guided wave behavior.

The Green's function for the problem is a self-consistent one for a driving current vector Dirac delta function applied at the guiding microstrip metal, $\mathbf{J} = j_x\delta(x-x_0)\hat{x} + j_z\delta(x-x_0)\hat{z}$ ($x_0 = 0$ for a centered strip). Although this Green's function is isotropic, it comes from a general spectral domain (Fourier transform) approach which is anisotropic. The Green's function is a dyadic, constructed as a 2×2 array relating tangential x- and z-components of surface current density to tangential electric field components. This Green's function is used to solve for the propagation constant. Determination of the field components is done in a second stage of processing, which in effect creates a large rectangular Green's function array, of 6×2 size in order to generate all electromagnetic field components, including those in the y-direction normal to the structure layers. The self-consistent problem is solved by expanding the surface currents on the guiding microstrip metal in an infinite expansion $J_x = \sum_{i=1}^{n_x} a_{xi} j_{xi}(x)$ and $J_z = \sum_{i=1}^{n_z} a_{zi} j_{zi}(x)$ and then requiring the determinant of the resulting system of equations to be zero. At this step of the problem, only the basis current functions need be provided and the complex propagation constant $\gamma = \alpha + j\beta$ is returned by the code. Of course, the summation limits $n_x$ and $n_z$ must be truncated at an appropriate value where convergence is acceptable.

Acquisition of the electromagnetic fields necessitates obtaining the basis function expansion coefficients $a_{xi}$ and $a_{zi}$, explicitly constructing the actual driving surface current density on the microstrip metal, finding the resulting top or bottom boundary fields, and then utilizing operators $P_{u,d}$ to pull up or down through the structure layers, generating the electric and magnetic fields in the process. Theory of $P_{u,d}$ operators is in [17] and their use in line plots in [18], the Cayley-Hamilton theorem in the Green's function for matrix exponentiation is in [19], and eigenvalue-eigenvector theory in the determination of $P_{u,d}$ with its matrix exponentiation is found in [17]. The entire solution method uses the constraint that the vertical side walls of the device are perfect electric walls, which can be shown to discretize the eigenvalues in the x-direction. These are the Fourier transform variables for the spectral domain, and an infinite set of them forms a complete set for the problem. Only a finite number of them are used, their maximum number being denoted by n.

To gain some idea of the trend of the propagation constant $\gamma = \alpha + i\beta$, a dispersion diagram is graphed in Fig. 1 (color; $\alpha$ is in dashed green, $\beta$ in solid blue) between 1 and 100 GHz for a device with air above the substrate and left-handed medium below with $Re[\varepsilon(\omega)] = -\varepsilon_r = -2.5$ and $Re[\mu(\omega)] = -\mu_r = -2.5$, substrate thickness $h_s = 0.5$ mm, microstrip width $w = 0.5$ mm, air region thickness $h_a = 5.0$ mm, and vertical wall separation $2b = 5.0$ mm. Also, $Im[\varepsilon(\omega)] = -\varepsilon_i = 0.0$ and $Im[\mu(\omega)] = -\mu_i = 0.0$ making the medium lossless (We also consider the microstrip metal lossless, although modifications for loss can be made [20].). There are two roots shown for even symmetry of the $J_z$ surface current component. They are mirror images of each other, with the correct sign of $\alpha$ in $\gamma$ in the region of evanescence from 6.5 GHz to 74 GHz ($\exp[-\gamma z]$ needs $\alpha > 0$ for forward propagation $z > 0$ when time dependence is $\exp\{i\omega t\}$ and $\beta > 0$). We identify these modes graphed as the fundamental modes since they exist even as $\omega \to 0$. Figure 1 inset (color) shows a blowup of the previous dispersion diagram between 1-6 GHz for these fundamental modes plus the next two higher order even symmetric modes. Out of the region of evanescence, $\alpha$ goes identically to zero, producing a pure propagating mode (only $\beta \neq 0$). These dispersion diagrams were produced with $n_x = n_z = 1$ although we have found $\gamma$ solutions at selected frequencies up to $n_x = n_z = 9$, the change in numerical value being in the fourth decimal place.

Propagation constant was determined to be $\gamma = (0.0000, \pm 2.22663)$ for the fundamental and $(0.0000, \pm 4.9556)$ for the first higher order mode at f = 5 GHz in the propagating region. In the evanescent region, the fundamental mode has $\gamma = (\pm 0.9394, \pm 2.1341)$ at 10 GHz. In the millimeter wave propagating region at 80 GHz, there are many solutions, and we list the two fundamentals and the next four higher order modes; $\gamma = (0.0000, \pm 1.7886), (0.0000, \pm 1.1777), (0.0000, \pm 0.90225), (0.0000, \pm 0.87369), (0.0000, \pm 0.69613),$ and $(0.0000, \pm 0.65065)$.

Figure 2 (color) shows a field line plot of the electric **E** (blue) and magnetic **H** fields (red) at 5.0 GHz for the fundamental mode with the line discretization set at about 0.02 mm. Physics of the field line directions is different for structures with left-handed media than for those with only regular media. This is very clearly evidenced in Fig. 2 for the propagation case. Maxwell's electric divergence equation in integral form is

$$\oint_S \mathbf{D}(x,y;t) \cdot d\mathbf{s} = \iint_A q(x,y;t) da = Q(t) \qquad (1)$$

in the cross-section A, d**s** an oriented line element, and da an area element. Since the constitutive relationship is $\mathbf{D} = \varepsilon \mathbf{E}$, (1) may be rewritten as

$$\int_{S_{top}} \mathbf{E}(x,y;t)/\varepsilon_{top} \cdot d\mathbf{s} + \int_{S_{LHM}} \mathbf{E}(x,y;t)/\varepsilon_{LHM} \cdot d\mathbf{s} = Q(t) \tag{2}$$

Taking into account that $\varepsilon_{LHM} = -|\varepsilon_{LHM}|$, this may be expressed as

$$\frac{1}{\varepsilon_{top}} \int_{S_{top}} \mathbf{E}(x,y;t) \cdot d\mathbf{s} - \frac{1}{|\varepsilon_{LHM}|} \int_{S_{LHM}} \mathbf{E}(x,y;t) \cdot d\mathbf{s} = Q(t) \tag{3}$$

Supposing charge to reside on the microstrip metal as a bilayer, negative on top and positive below, contour contributions to the integral in (1) will oppose each other in (3) if the integral $\int_S \mathbf{E}(x,y;t) \cdot d\mathbf{s}$ keeps the same negative sign in going from above the interface to below it into the left-handed medium. This is precisely what occurs, as **E** field lines point downward above and upward below the microstrip. The enclosing surface S may be imagined as a closed line encircling the microstrip, with its normal pointing outward, the direction of the differential element d**s**. By extending the $S_{top}$ surface contour along the x-axis exactly through the center of the metal, making it closed, and doing the same for the $S_{LHM}$ surface contour, these integrations being opposite and through $\mathbf{E} = 0$, form representations of the upper or lower components of the bilayer charges. The particular charge distribution seen on the metal, in terms of both its infinitesimal y separation and x-distribution (the main contribution produces a z-directed symmetric "u-shaped" $J_z$) will vary of course, depending on the particular mode examined.

Away from the driving currents on the microstrip metal,

$$D_{n,top} = D_{n,LHM} \quad ; \quad \mathbf{n} \times \left[\mathbf{E}_{top} - \mathbf{E}_{LHM}\right] = 0 \tag{4a,b}$$

Looking at the electric field lines, and using the constitutive relationship, (4ab) can be recast as

$$\varepsilon_{top} E_{n,top} = \varepsilon_{LHM} E_{n,\, LHM} \quad ; \quad E_{tan,top} = E_{tan,\, LHM} \qquad (5a,b)$$

Inserting $\varepsilon_{LHM} = -|\varepsilon_{LHM}|$ into (4a) gives

$$\varepsilon_{top} E_{n,top} = -|\varepsilon_{LHM}| E_{n,\, LHM} \qquad (6)$$

Invoking (5b) and (6) at the interface for $b > |x| > w/2$ means that the electric field lines above and below the interface must both point toward or away from the interface, which is seen in Fig. 3.

Maxwell's magnetic curl equation in integral form is

$$\oint_L \mathbf{H}(x,y;t) \cdot d\mathbf{l} = \iint_A \mathbf{J}(x,y;t) \cdot d\mathbf{a} = I(t) \qquad (7)$$

where the current density **J** is made up of displacement, volumetric and microstrip surface currents. The line integral on the left may be broken down into its pieces like (2) above, giving

$$\int_{L_{top}} \mathbf{H}(x,y;t) \cdot d\mathbf{l} + \int_{L_{LHM}} \mathbf{H}(x,y;t) \cdot d\mathbf{l} = I(t) \qquad (8)$$

Using the constitutive relationship $\mathbf{B} = \mu \mathbf{H}$ and $\mu_{LHM} = -|\mu_{LHM}|$,

$$\frac{1}{\mu_{top}} \int_{L_{top}} \mathbf{B}(x,y;t) \cdot d\mathbf{l} - \frac{1}{|\mu_{LHM}|} \int_{L_{LHM}} \mathbf{B}(x,y;t) \cdot d\mathbf{l} = I(t) \qquad (9)$$

This has exactly the same form as (3), and we may wonder if here as well the contour pieces have some relationship to each other in terms of their polarity. They do and this is discovered if we examine the continuity condition for magnetic field **B** at the ordinary dielectric/left-handed medium interface located at, say some $x < 0$ off of the microstrip metal, that is for $b < x < -w/2$.

Normal **B** field components must be continuous across the interface, whereas tangential **H** field components must be discontinuous by the surface current **J**$_s$. Thus,

$$B_{n,top} = B_{n,\ LHM} \quad ; \quad \mathbf{n} \times [\mathbf{H}_{top} - \mathbf{H}_{LHM}] = \mathbf{J}_s \tag{10a,b}$$

Looking at the magnetic field lines away from the microstrip, **J**$_s$ = 0, and again using the constitutive relationship, (10ab) can be recast as

$$\mu_{top} H_{n,top} = \mu_{LHM} H_{n,\ LHM} \quad ; \quad H_{\tan,top} = H_{\tan,\ LHM} \tag{11a,b}$$

Inserting $\mu_{LHM} = -|\mu_{LHM}|$ into (11a) gives

$$\mu_{top} H_{n,top} = -|\mu_{LHM}| H_{n,\ LHM} \tag{12}$$

Applying (11b) and (12) at the interface for b > |x| > w/2 means that the magnetic field lines above and below the interface must both point toward or away from the interface. This is indeed the case as seen in Fig. 2. And it is found that integral $\int_L \mathbf{B}(x,y;t) \cdot d\mathbf{l}$ holds the same positive sign in going from above the interface to below it into the left-handed medium.

A different visualization technique is provided in Figs. 3 (color) and 4 (color) to complete the field assessment where we have produced color variation of the magnitude of the electric E = sqrt[$\sum_{i=1}^{3} E_i^2$] and magnetic H = sqrt[$\sum_{i=1}^{3} H_i^2$] fields with overlays of the **E** and **H** vectors with arrows sized according to magnitude. Figs 3 and 4 are produced from a grid of 8372 points stored in a 1 MB file. Finally, power flow is given by the Poynting vector **P** = **E**×**H**. Assessment of **P** using only the cross-sectional fields yields information about the propagation in the guide direction, $\hat{z}$, so $\mathbf{P}_{guide} = \mathbf{P} \cdot \hat{z} = \mathbf{E}_t \times \mathbf{H}_t = P_{guide}\hat{z}$. Fig. 5 (color) gives $P_{guide}$ in a distribution, and also shows the regions where a reversal from the dominant positive flow occurs. The maximum deepness of the reversal is 18.1 %. Of course, total power $P_T$ down the guiding structure is given by $P_{Tz} = \oiint_S \mathbf{P} \cdot d\mathbf{a} = \oiint_S \mathbf{E} \times \mathbf{H} \cdot d\mathbf{a}$.

Another important question is causality. Although the calculations presented here and in [21] have taken the LHM substrate loss as zero, we have performed additional simulations with loss in the LHM by letting $\varepsilon_i = -0.0025$ ($\text{Im}[\varepsilon(\omega)] = -\varepsilon_i > 0$) at f = 5 GHz. This leads to causal waves which decay in either the ± z directions for, respectively, $\alpha, \beta > 0$ or $\alpha, \beta < 0$. Change of the γ eigenvalue due to the added loss is small, with no apparent change in β, $\alpha/k_0$ rising from 0 to 0.002062. Current expansion coefficient change is from $[a_{x1}, a_{z1}] = [(0, -1.2043 \times 10^{-3}), (1,0)]$ to $[(-4.7273 \times 10^{-7}, -1.2043 \times 10^{-3}), (1,0)]$ and the field components change in the 4$^{th}$ decimal place or higher, causing no visual alteration of the field distributions.

In conclusion, demonstrations of guided wave propagation (and non-propagation) down a guided wave left-handed structure has been elucidated in terms of the new physics. Completely new propagation constant behavior and field arrangements suggest the possibility of entirely different types of devices for future electronics compatible with integrated circuit and solid state technology.

M. Daniel of DCS Corp. is thanked for his software contributions. Encouragement of Dr. Gerald M. Borsuk of ESTD at NRL is acknowledged, and Dr. W. J. Moore is thanked for his interest in this subject.

FIGURE CAPTIONS (all in color)

1. Complex propagation constant γ versus frequency f over the range 1-100 GHz. Fundamental modes for the microstrip configuration with a left-handed substrate.

   Inset: Lower end mode region 1-6 GHz.

2. Electromagnetic field line plot showing electric **E** (blue) and magnetic **H** (red) fields at 5.0 GHz, in the lower propagating mode region.

3. Electromagnetic field color plot showing electric field magnitude E (color variation) with an overlaid vector field **E** (black arrows) at 5.0 GHz.

4. Electromagnetic field color plot showing electric field magnitude H (color variation) with an overlaid vector field **H** (black arrows) at 5.0 GHz.

5. Electromagnetic color plot showing the Poynting vector $P_{guide}$ (color variation) in the guiding z direction at 5.0 GHz.

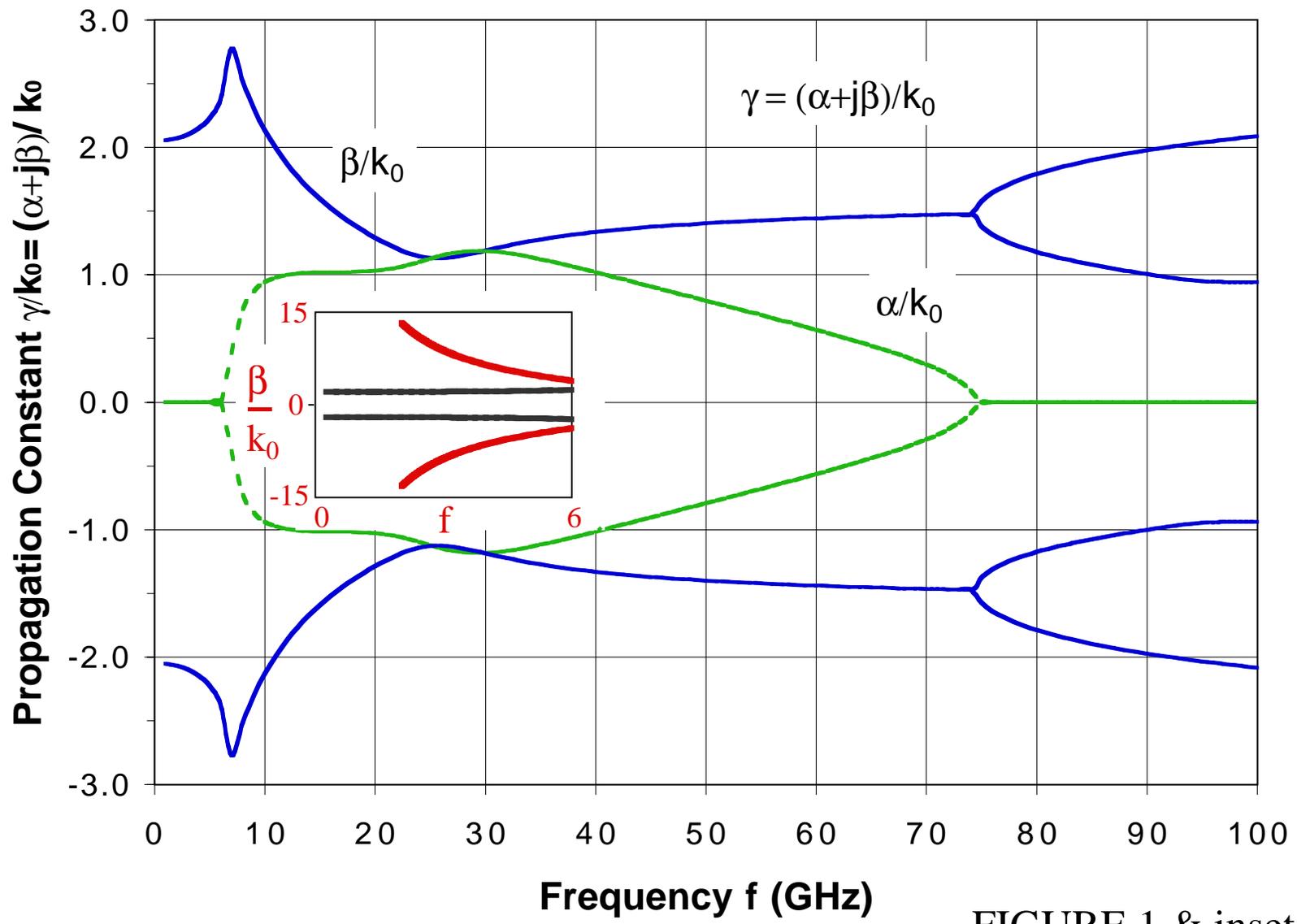

FIGURE 1 & inset

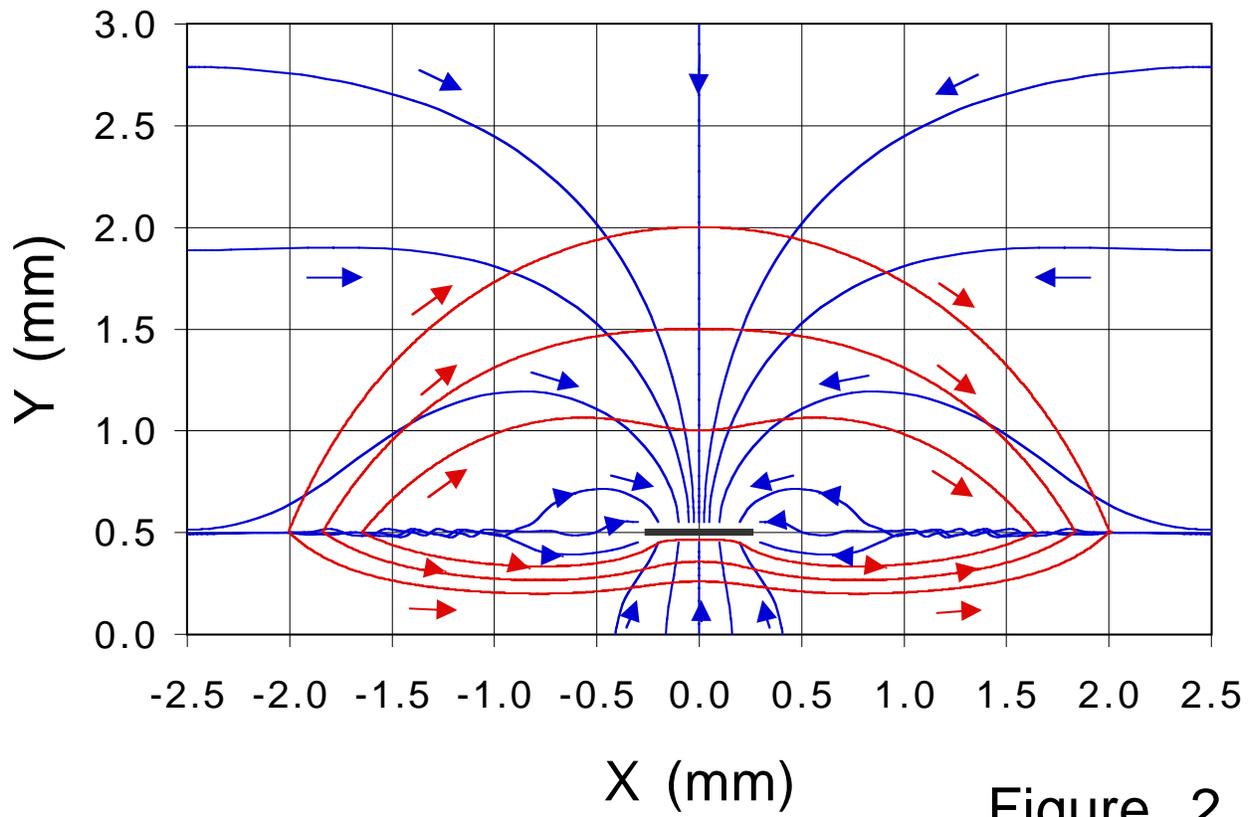

Figure 2

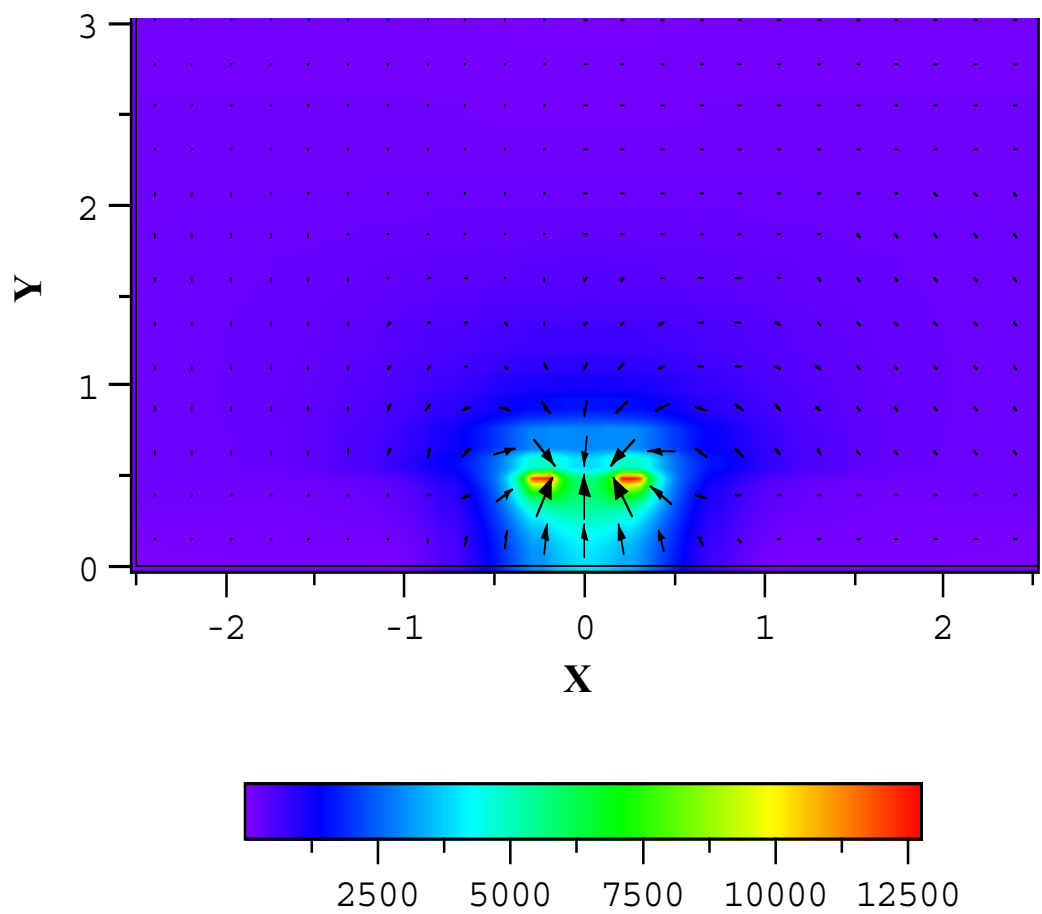

Figure 3

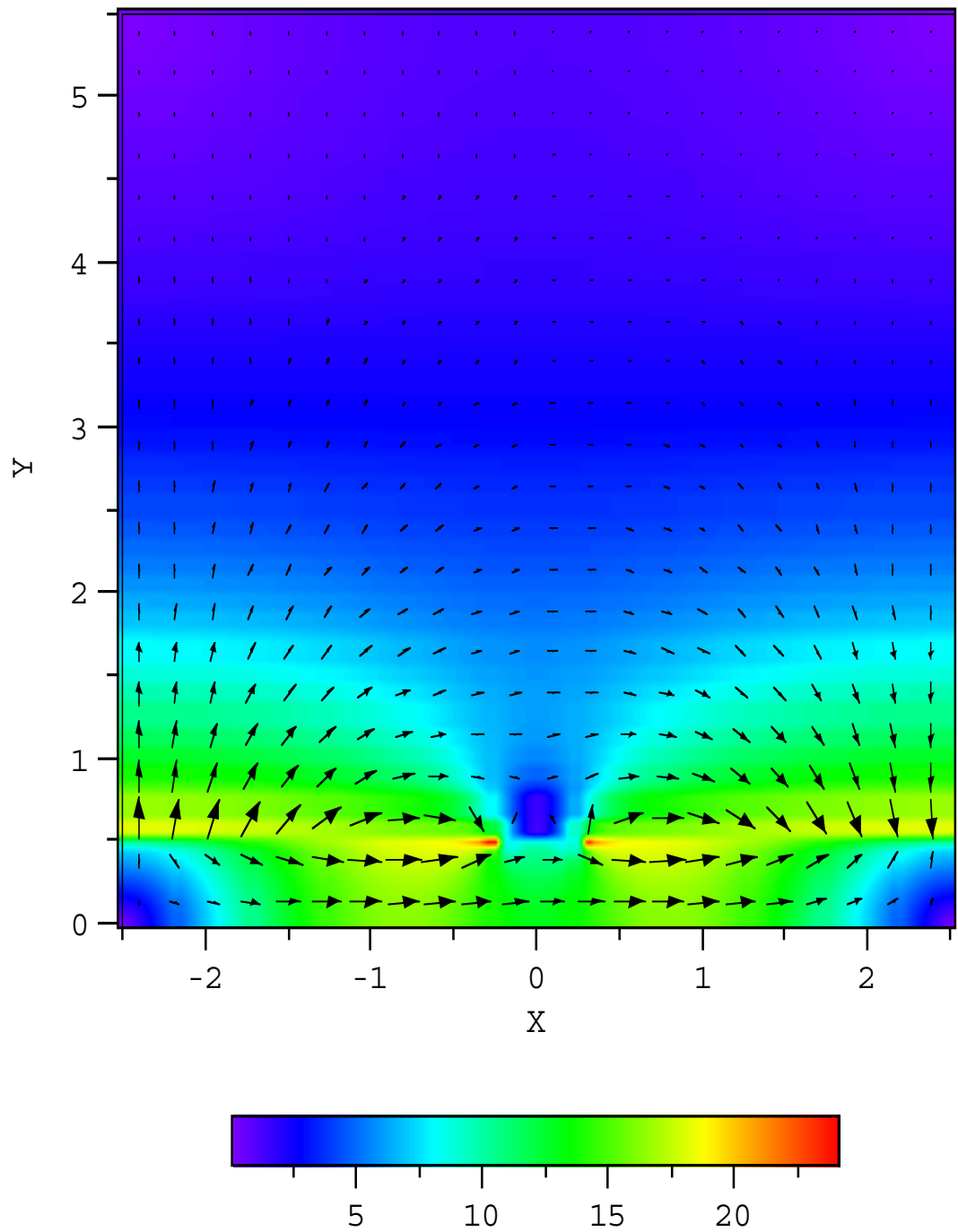

Figure 4

Figure 5

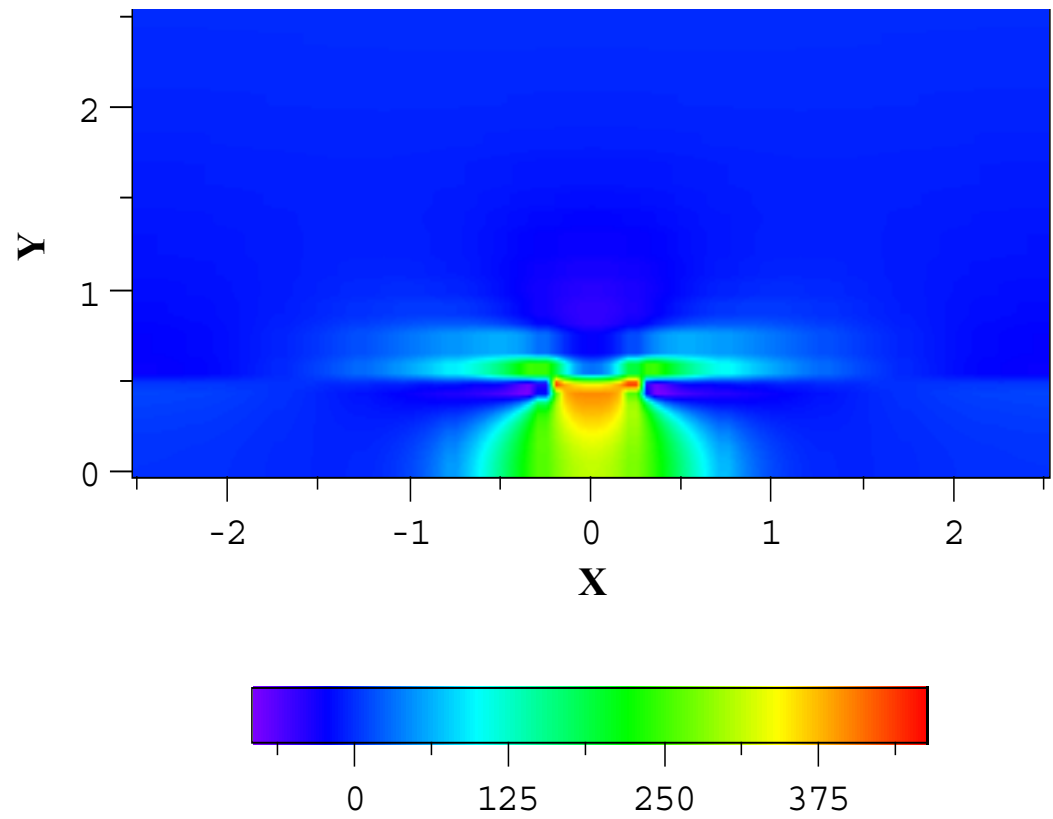